\documentclass[twocolumn,pra,aps,superscriptaddress]{revtex4-1}
\usepackage[latin9]{inputenc}
\usepackage{mathtools}
\usepackage{amsbsy}
\usepackage{amssymb}
\usepackage{amstext}
\usepackage[unicode=true,pdfusetitle,
 bookmarks=false,
 breaklinks=false,pdfborder={0 0 1},backref=false,colorlinks=false]
 {hyperref}
\hypersetup{
 bookmarksnumbered=false,bookmarksopen=false}
\makeatletter

\@ifundefined{textcolor}{}
{%
 \definecolor{BLACK}{gray}{0}
 \definecolor{WHITE}{gray}{1}
 \definecolor{RED}{rgb}{1,0,0}
 \definecolor{GREEN}{rgb}{0,1,0}
 \definecolor{BLUE}{rgb}{0,0,1}
 \definecolor{CYAN}{cmyk}{1,0,0,0}
 \definecolor{MAGENTA}{cmyk}{0,1,0,0}
 \definecolor{YELLOW}{cmyk}{0,0,1,0}
}

\newcommand{\beq}{\begin{equation}}
\newcommand{\eeq}{\end{equation}}
\newcommand{\beqa}{\begin{eqnarray}}
\newcommand{\eeqa}{\end{eqnarray}}

\usepackage{amsmath}
\usepackage{graphicx}
\usepackage{amssymb}
\usepackage{txfonts,color}
\makeatother

\begin{document}
\title{Robust Detection of High-Frequency Signals at the Nanoscale}
\author{Carlos Munuera-Javaloy}
\affiliation{Department of Physical Chemistry, University of the Basque Country UPV/EHU, Apartado 644, 48080 Bilbao, Spain}
\author{Yue Ban}
\affiliation{Department of Physical Chemistry, University of the Basque Country UPV/EHU, Apartado 644, 48080 Bilbao, Spain}
\affiliation{College of Materials Science and Engineering, Shanghai University, 200444
	Shanghai, China}
\author{Xi Chen}
\email{xchen@shu.edu.cn} 
\affiliation{Department of Physical Chemistry, University of the Basque Country UPV/EHU, Apartado 644, 48080 Bilbao, Spain}
\affiliation{International Center of Quantum Artificial Intelligence for Science and Technology~(QuArtist)
and Physics Department, Shanghai University, 200444 Shanghai, China}
\author{Jorge Casanova}
\email{jcasanovamar@gmail.com} 
\affiliation{Department of Physical Chemistry, University of the Basque Country UPV/EHU, Apartado 644, 48080 Bilbao, Spain}
\affiliation{IKERBASQUE,  Basque  Foundation  for  Science,  Plaza Euskadi 5,  48009  Bilbao,  Spain}

\begin{abstract} 
We present a method relying on shortcuts to adiabaticity to achieve quantum detection of high frequency signals at the nanoscale in a robust manner. More specifically, our protocol delivers tailored amplitudes and frequencies for control  fields that, firstly, enable the coupling of the sensor with high-frequency signals and, secondly, minimise errors that would otherwise spoil the detection process. To exemplify the method, we particularise to detection of signals emitted by fast-rotating nuclear spins with nitrogen vacancy center quantum sensors. However, our protocol is  straightforwardly applicable to other quantum devices such as silicon vacancy centers, germanium vacancy centers, or divacancies in silicon carbide.
 \end{abstract}
\maketitle

\section{Introduction} Nanoscale nuclear magnetic resonance (Nanoscale NMR) is a flourishing research area leading to detection and control of magnetically active nuclear spin species with unprecedented spatial resolution~\cite{Mamin13, Muller14, DeVience15, Degen17, Schwartz19, Holzgrafe20, Zopes18, Zopes18bis, Bradley19, Abobeih19}.  This ability has profound applications in different contexts such as the narrowband measurement of electromagnetic fields~\cite{Schmitt17, Boss17, Glenn18}, the detection of fluids in nanoscale sized voxels~\cite{Staudacher13}, single molecule spectroscopy~\cite{Shi15, Lovchinsky16, Aslam17},  or in-cell thermometry~\cite{Kucsko13, Choi20}. All these applications rely on the presence of controllable minute-sized quantum sensors that play the role of macroscopic detection coils in standard NMR apparatus~\cite{Glover02}. Among currently available quantum sensors we can mention, e.g., silicon vacancy centers~\cite{Rogers14}, divacancies in silicon carbide~\cite{Christle17}, germanium vacancy centers~\cite{Siyushev17}, and nitrogen vacancy (NV) centers~\cite{Doherty13}. 

In particular, the NV center in diamond has been extensively studied owing to its excellent properties for nanoscale NMR tasks~\cite{Dobrovitski13, Rondin14, Schirhagl14, Wu16}. Namely, the electron spin of the NV center exhibits quantum coherence at ambient conditions~\cite{Balasubramanian08}, thus it enables spectroscopy of biomolecules in their natural environment~\cite{Shi15, Aslam17}. In addition, the possibility of delivering shallow NVs~\cite{Pham16} enables detection of samples on the diamond surface~\cite{Muller14, Kehayias17}, whilst NV centers embedded in nanodiamonds can be used as nanosensors {\it in vivo}  as a consequence of their excellent biocompatibility~\cite{Kehayias17, Chipaux18}. In this context, extending the quantum coherence of the NV is crucial, as this permits a larger interrogation time with the target and isolation from environmental noise. In the case of NVs, this is met by dynamical decoupling (DD) techniques in the form of pulsed~\cite{Maudsley86, Uhrig08, Pasini08, Souza11, Wang11, Souza12, Casanova15, Wang16, Lang17, Wang19} or continuous~\cite{Hirose12, Cai13, Puebla18, Arrazola19} microwave (MW) sequences.  

Particularly interesting for nanoscale NMR is the regime of strong static magnetic fields~\cite{Aslam17}. In this scenario, thermal spin polarisation of target samples get increased leading to a larger NMR signal contrast~\cite{Levitt08}, nuclear and electron spins exhibit long coherence times~\cite{Reynhardt01}, and structural parameters such as the chemical shift get increased~\cite{Levitt08}. As a counterpart, the spin of nuclei rapidly precess at strong magnetic fields. This challenges their identification as the Hartmann-Hahn resonance condition~\cite{Hartmann62}, which is a decisive requisite for quantum detection,  cannot be satisfied with realistic MW power. To circumvent this problem, it was recently proposed the delivery of extended MW pulses with a modulated amplitude~\cite{Casanova18}. These pulses imprint on the NV spin evolution high frequencies that meet those of nuclear spins. However, these schemes are only valid in conditions involving low errors on the controls. Other schemes, such as adiabatic chirped pulses~\cite{Genov19}, present an excellent robustness. However, as pointed out by the authors in~\cite{Genov19}, this resilience significantly decays at large static magnetic fields. In this manner, the design of DD sequences that stabilise the sensor under large control errors, whilst enable the coupling with high frequency signals (such as those emitted by fast rotating nuclear spins, i.e. at strong magnetic fields) is of clear importance owing to the potential advantages of nanoscale NMR in this regime.

In this Letter, we present a method that achieves nanoscale NMR at strong magnetic fields in realistic conditions that involve large errors on the controls. To this end, we integrate shortcuts to adibaticity (STA) techniques~\cite{Erik13,Guery19} in the design of the DD sequences that drives the interaction between  sensor and  target signals. By means of detailed numerical simulations, we demonstrate that our protocol enables resilient quantum magnetometry in relevant nanoscale NMR scenarios such as the detection of nearby nuclear spins, as well as of nuclear clusters at strong magnetic fields. We exemplify our theory in NV centers in diamond, but this is general and applicable to other solid-state sensors.

\section{The model} We consider a Hamiltonian that describes an NV center coupled to a target signal, and driven by a MW field. This is 
\begin{equation}\label{start}
H = DS_z^2 - \gamma_e B_z S_z + H_T + \sqrt{2} S_x \Omega(t) \cos{[\omega t - \Delta(t) - \phi]}.
\end{equation}
Here $D=(2\pi)\times 2.87$ GHz is the zero-field splitting, and $\gamma_e = (2\pi)\times 28.024$ GHz/T is the electronic gyromagnetic ratio. The magnetic field $B_z$  is aligned with the NV axis, $S_{z, x}$ are spin-1 matrices of the NV center, and $H_T$ denotes the coupling of the NV with the target signal.  For instance, $H_T = -\gamma_N B_z I_z  +S_z  \vec{A}\cdot \vec{I}$ in case of having a nearby nuclear spin, with $\gamma_N$ being the nuclear gyromagnetic ratio, $\vec{A}$ the hyperfine vector that couples the NV and the nucleus, and $\vec{I}$ is the spin operator of the nucleus. On the other hand, when considering a classical signal that models, e.g., a nuclear spin cluster out of the diamond lattice~\cite{Aslam17, Laraoui11} we may have $H_T = \Gamma S_z \cos(\omega_s t)$.  The last term in Eq.~(\ref{start}), i.e. the MW control term, encompasses the functions $\Omega(t)$ and $\Delta(t)$ that our method will set such that they lead to optimal detection of targets at strong magnetic fields under severe error conditions. 

The dynamics associated to Eq.~(\ref{start}) can be analysed in the following picture: Firstly, we move to a rotating frame with respect to (w.r.t.) $H_0 = DS_z^2 - \gamma_e B_z S_z$ and set the MW field frequency as $\omega = D + |\gamma_e|B_z$. This leads to the Hamiltonian 
$H =  H_T + \frac{\Omega(t)}{2} \left[|1\rangle\langle 0| e^{i\Delta(t)} e^{i\phi} + \rm{H.c.}\right]$ where the terms involving transitions to the $|-1\rangle$ spin state of the NV have been neglected by invoking the rotating wave approximation (RWA). Finally, we move to a second  rotating frame w.r.t. $\frac{-\delta(t) }{2} \sigma_z$, where $\delta(t)$ is  $\int_{t_0}^t \delta(s) ds = \Delta(t)$, and  $\sigma_z = |1\rangle\langle 1| -|0\rangle\langle 0|$. In this manner, the Hamiltonian of the system reads 
\begin{equation}\label{general}
H = H_T  + \frac{\Omega(t)}{2} \sigma_{\phi} + \frac{\delta(t) }{2} \sigma_z,  
\end{equation}
with $\sigma_\phi = |1\rangle\langle 0| e^{i\phi} +  |0\rangle\langle 1| e^{-i\phi}$. Note that $\phi=0$ $(-\pi/2)$ implies $\sigma_{\phi} = \sigma_x$ $(\sigma_y)$.

The control term $H_c= \frac{\Omega(t)}{2} \sigma_{\phi} + \frac{\delta(t) }{2} \sigma_z$ causes periodic population exchanges in the $[|1\rangle, |0\rangle]$ spin manifold, thus it imprints in the NV dynamics a set of frequencies. Ultimately, by tuning the periodicity of these spin exchanges one would get a resonant interaction between the NV and the target signal leading to quantum detection. An archetypical example of the latter is the HH resonance that reduces to $\Omega \approx \gamma_N B_z - \frac{1}{2} A_Z$ when $H_T = -\gamma_N B_z I_z  +S_z  \vec{A}\cdot \vec{I}$, this is in the presence of a single nearby nuclear spin. Also, if the target is a classical signal, i.e. $H_T = \Gamma S_z \cos(\omega_s t)$, the HH condition is $\Omega = \omega_s$. We note that the achievement of the HH condition is challenging at strong static magnetic fields as $\Omega$ is proportional to $B_z$, which implies that high MW power should be delivered to the sample.

Other schemes involving extended $\pi$ pulses have been proposed in the literature to achieve couplings with rapidly oscillating signals~\cite{Casanova18}. However, these extended $\pi$ pulses suffer from control errors which seriously limits their performance in realistic scenarios. We will later demonstrate this with specific numerical simulations  at $B_z = 3$ T. To overcome this challenge we integrate STA techniques in the design of $\pi$ pulses in such a way that they enable the coupling with targets at strong magnetic fields in a robust manner. 

\section{The method} Inspired by the concept of STA \cite{Erik13,Guery19}, we parameterize the NV spin state evolution as \cite{Daems13} 
\begin{equation}
\label{wavefuction}
|\phi (t) \rangle = \left[\cos\left(\frac{\theta}{2}\right) e^{i\frac{\beta}{2}} |1\rangle + \sin \left(\frac{\theta}{2}\right) e^{-i\frac{\beta}{2}} |0\rangle \right] e^{i \gamma},
\end{equation}
with $\theta\equiv\theta(t)$ and $\beta\equiv\beta(t)$ being  the polar and azimuthal angles on the Bloch sphere, and
a phase $\gamma\equiv\gamma(t)$. When inserting Eq. (\ref{wavefuction}) into
the time-dependent Schr\"{o}dinger equation governed by the control Hamiltonian $H_c$, we get the next auxiliary equations 
\beqa
\label{dottheta}
\dot{\theta} &=& \Omega(t)\sin(\beta),\label{InvCondition1}	
\\
\label{dotbeta}
\dot{\beta} &=&  \dot{\theta} \cot(\theta)\cot(\beta)-\delta(t),
\\
\dot{\gamma} &=& - \dot{\theta}\cot(\beta)/(2 \sin (\theta))\label{dotgamma}.
\eeqa
For the sake of simplicity, in the previous equations we have particularised to the case $\sigma_\phi = \sigma_x$, but the formalism is equally applicable to  $\sigma_\phi$. Equations~(\ref{dottheta}, \ref{dotbeta}, \ref{dotgamma})  connect the Rabi frequency $\Omega(t)$ and the detuning $\delta(t)$ with the $\theta$ and $\beta$ angles. 
Note that, similar expressions to Eqs.~(\ref{dottheta}, \ref{dotbeta}, \ref{dotgamma}) can be derived from a dynamical invariant \cite{Lu13,Ruschhaupt12}, as well as by inverting the Madelung representation~\cite{Qi17}.  

To achieve a $\pi$ pulse, e.g. from $|1\rangle$ at $t=0$ to $|0\rangle$ at $t=t_\pi$, one has to impose the following  boundary conditions to the wavefunction in Eq~(\ref{wavefuction}),
\begin{equation}
\theta(0) = 0, ~ \theta(t_\pi) = \pi.
\label{Boundary}
\end{equation}
A possible parametrisation for $\theta$ and $\beta$ is: $\theta= \pi t/t_\pi$ and $\beta= \pi/2$ leading to $\Omega(t) = \pi/t_\pi$ and $\delta(t) = 0$. Then, we would get a top-hat $\pi$ pulse at $t_{\pi}$ (note that $\int_{0}^{t_\pi} \Omega(t)dt=\pi$). On the other hand, we note that there exists much freedom to tailor the functions $\theta$ and $\beta$, such that one gets pulse designs that enable the coupling with fast precessing nuclei and, at the same time, they are resilient to severe control errors.

Regarding the coupling of the NV  with rapidly oscillating signals, one can demonstrate that maximal NV-target interaction strength is achieved if the following {\it coupling condition} holds~(for details regarding the derivation of the coupling condition see Appendix A)
\begin{equation}
\label{CouplingCondition}
\int_{0}^{t_\pi}\cos (\theta) \cos(k\omega_m t)dt = 0.
\end{equation}
Here, $\omega_m = 2\pi/T$ with $T$ being the period of the employed DD sequence, and $k \in \mathbb{N} $ labels  the harmonic that will carry the NV-target  coupling. Hence, Eq.~(\ref{CouplingCondition}) represents a 
first requisite for the $\theta$ function.

\begin{figure*}[t]
\hspace{-0. cm}\includegraphics[width=1.5\columnwidth]{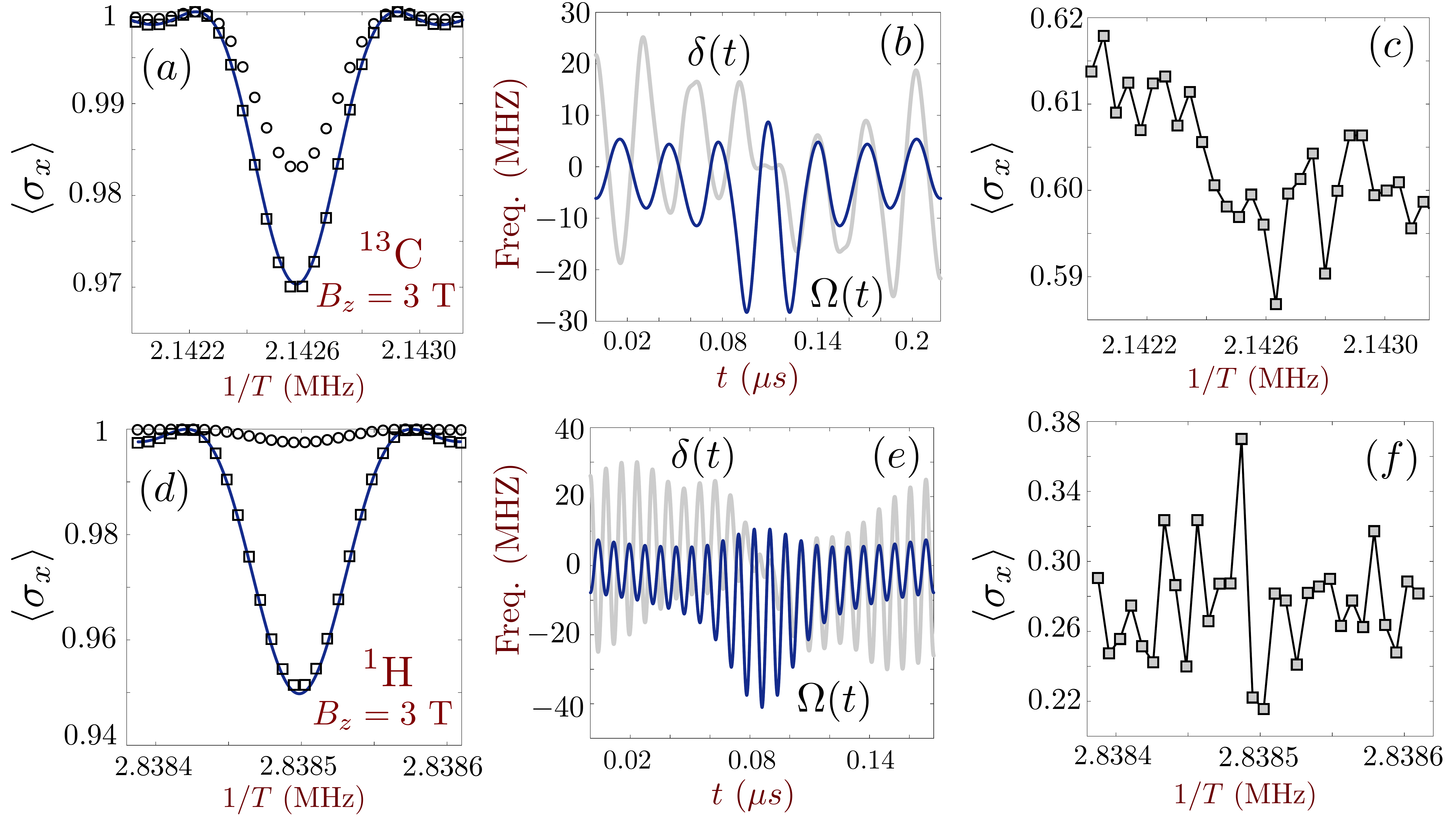}
\caption{Distinct Nanoscale NMR scenarios at a strong magnetic field $B_z = 3$ T involving a $^{13}$C nucleus in (a), (b), and (c), and a $^{1}$H cluster in (d), (e), and (f). In (a) we show the Nanoscale NMR spectrum, i.e. the $\langle\sigma_x\rangle$ of the NV, in three different situations:  Solid-blue curve corresponds to an ideal escenario involving instantaneous $\pi$ pulses. The squares represent the spectrum obtained with our method, while circles uses standard top-hat pulses. In all cases we have repeated the XY8 sequence 102 times. This implies that 816 $\pi$ pulses have been employed leading to a final sequence time $\approx 0.19$ ms. In (b) we show the controls $\Omega(t)$ (dark curve) and $\delta(t)$ (clear curve) used in (a) for computing the spectrum including squares. Note $\Omega(t)$ and  $\delta(t)$ are in units of frequency, while the $\pi$ pulse induced by these controls has a duration $\approx 0.22\ \mu$s. (c)  Obtained spectrum with the extended $\pi$ pulses in Ref.~\cite{Casanova18}. (d) Similar to (a) but having a cluster of $^1$H nuclei as a target (see main text). (e) Controls used for finding the spectrum (squares) in (d), in this case leading to a $\pi$ pulse duration of $\approx 0.17\ \mu$s. (f) Similar to (c) but applied to $^1$H nuclei (see main text). In (d), (e), and (f) we repeat the XY8 sequence 102 times, which in this case leads to a final sequence time $\approx 0.14$ ms.}
\label{results}
\end{figure*}
Further constraints have to be imposed in the dynamics of $|\phi (t) \rangle$ to cancel  control errors during quantum detection. Typically, these errors are: $(i)$ Deviations in the Rabi frequency, i.e. $\Omega(t)  \rightarrow\Omega(t)(1+\xi_\Omega)$, as a consequence of MW power variations denoted by $\xi_\Omega$. And, $(ii)$, errors in the $\delta(t)$ function (i.e. $\delta(t) \rightarrow \delta(t)+\xi_\delta$) with $\xi_\delta$ being a frequency offset that appears owing to, e.g., undetermined stress conditions in the diamond and/or because of nearby electronic impurities leading to NV energy shifts. In this scenario we use perturbation theory over  $|\phi (t) \rangle$ during the $\pi$ pulse, and calculate the transition probability $P(t_\pi)$ of having an NV spin-flip driven by an imperfect $\pi$ pulse (up to second order in $\xi_\Omega$ and $\xi_\delta$). This reads
$P(t_\pi) \approx 1-\frac{1}{4}\left|\int_{0}^{t_\pi}dt e^{i 2 \gamma}\left(\xi_\Delta\sin(\theta) - i2\xi_\Omega\dot{\theta}\sin^2(\theta)\right)\right|^2$. For more details regarding the derivation of $P(t_\pi)$ see Appendix B. In this manner, the second requisite for $\theta$ and $\gamma$ is the {\it error cancelation condition} that eliminates control errors during the NV spin-flip. This reads
\begin{equation}\label{errorcancelation}
\left|\int_{0}^{t_\pi}dt e^{i 2 \gamma}\left(\xi_\Delta\sin(\theta) - i2\xi_\Omega\dot{\theta}\sin^2(\theta)\right)\right| = 0.
\end{equation}

Once we get expressions for $\theta$ and $\gamma$ (and consequently to $\beta$ as Eq.~(\ref{dotbeta}) relates $\beta$ with $\theta$ and $\gamma$) one can find the control parameters $\Omega(t)$ and $\delta(t)$ by solving Equations~(\ref{dottheta}, \ref{dotbeta}).

In order to interpolate a function for $\theta$, we use an ansatz inspired by the Blackman function~\cite{Blackman58}. This is 
\begin{equation}
\label{theta}
\theta(t) = \alpha_0 + \alpha_1 \cos\left(\frac{\pi}{t_\pi}t\right)+\alpha\sin\left(\frac{2\pi\lambda}{t_\pi} t\right),
\end{equation}
where $\lambda$ is a free tunable parameter that regulates the $\pi$ pulse length as $t_\pi = \lambda T/k$, see Appendix A.  In addition, $\alpha_0$, $\alpha_1 $, and $\alpha$ are parameters that we will adjust to hold the previously commented conditions. In particular, when the boundaries in Eq.~(\ref{Boundary}) are applied to $\theta(t)$, we get $\alpha_0=-\alpha_1=\pi/2$, while the additional parameter $\alpha$ will be selected to fulfill the coupling condition in Eq.~(\ref{CouplingCondition}).

Now, we pose the following ansatz for $\gamma(t)$
\begin{equation}
\label{gamma}
\gamma(t) = \theta + \eta_1\sin(2\theta)+\eta_2\sin(4\theta),
\end{equation}
that introduces two additional free parameters $\eta_{1}$ and $\eta_{2}$. The expression for $\gamma(t)$ can be combined with Eq.~(\ref{dotgamma}) leading to
\begin{equation}
\label{beta}
\beta =\cos^{-1}\left(\frac{-2M\sin(\theta)}{\sqrt{1+4M^2\sin^2(\theta)}}\right), 
\end{equation}
where $M = 1+2\eta_1\cos(2\theta)+4\eta_2\cos(4\theta)$. We will use $\eta_1$ and $\eta_2$ to achieve  Eq.~(\ref{errorcancelation}) over some reasonable error interval. In this manner, undesired NV transitions caused  by  errors in the Rabi frequency and detuning get cancelled. This assures reliable detection of nuclear spins at large magnetic fields and under realistic conditions as it is shown in the following section.

\section{Numerical results} We demonstrate the performance of our method with detailed numerical simulations in relevant Nanoscale NMR escenarios. In particular, we have computed the evolution of an NV under an XY8 sequence in the presence of a nearby $^{13}$C nuclear spin, as well as under the influence of a classical electromagnetic wave  modelling a $^1$H nuclear spin cluster. In both cases we consider a strong  magnetic field $B_z = 3$~T~\cite{Aslam17}. We compare the obtained Nanoscale NMR spectra  in situations involving: Standard top-hat $\pi$ pulses, extended  $\pi$ pulses that follow the scheme in~\cite{Casanova18}, and $\pi$ pulses designed with our method that incorporates STA techniques. 

The results are presented in Fig.~\ref{results}. In (a) we show the computed spectra (encoded in the expectation value  $\langle\sigma_x\rangle$ of the NV center) of a problem involving an NV coupled to a nearby $^{13}$C nucleus (then $H_T = -\gamma_N B_z I_z  +S_z  \vec{A}\cdot \vec{I}$). The nucleus is at a distance of $1.1$ nm from the NV, such that its hyperfine vector $\vec{A} = (2\pi)\times[-4.81, -8.331, -26.744]$ KHz. The solid-blue line corresponds to the spectrum that would appear if instantaneous pulses (this is $\pi$ pulses with infinite MW energy) were delivered to the system. In addition, this solid-blue line has been obtained without introducing control errors. Then, this constitutes an idealised experimental scenario. 
The spectrum represented by the squares in Fig.~\ref{results} (a) has been calculated by using our method based on STA techniques. The particular values for the control parameters $\Omega(t)$ and $\delta(t)$ are shown in Fig.~\ref{results} (b), and have led to a $\pi$ pulse of length $t_{\pi} = 0.21 \ \mu$s. In addition, the reader can find an animation of the trajectory in the Bloch sphere of the NV spin induced by $\Omega(t)$ and $\delta(t)$ in~\cite{SM}.  We want to remark that, a detuning error of $\xi_\delta = (2\pi)\times1$ MHz, as well as a Rabi frequency deviation of $\xi_\Omega = 0.5\%$ are included in our numerical simulations. Even in these conditions involving significant errors, the spectrum produced by our method (squares) overlaps well with the ideal one (solid-blue). On the other hand, the spectrum represented by circles in Fig.~\ref{results} (a) has been computed with standard top-hat $\pi$ pulses with a Rabi frequency ($\Omega_{\rm th}$) that equals the maximum of $\Omega(t)$ in our method, see Fig.~\ref{results} (b). More specifically, this is $\Omega_{\rm th} \approx (2\pi) \times 30$ MHz). It is noteworthy to mention that the spectral contrast achieved by top-hat pulses (this is the peak depth of the spectrum with circles) is significantly lower than the one achieved by our method, which demonstrates the better performance of the latter. In Fig.~\ref{results} (c) we show the spectrum computed with the extended pulses in Ref.~\cite{Casanova18} which include the same errors on the controls ($\xi_\delta = (2\pi)\times1$ MHz, and $\xi_\Omega = 0.5\%$). Notably, the extended pulses in Ref.~\cite{Casanova18} produce a completely distorted spectrum that does not allow to identify the resonance of the $^{13}$C. As a further comment, in absence of control errors our method and the one in Ref.~\cite{Casanova18} lead to similar results. However, under the presence of significant error sources our protocol is clearly superior. 

In Fig.~\ref{results} (d) we present the spectra that result of averaging the response of several NVs, each of them with a different detuning error, whilst they are all coupled to the same classical electromagnetic wave. Thus, $H_T = \Gamma S_z \cos(\omega_s t)$, where we employ $\Gamma = (2\pi)\times 28$ kHz in the simulations. This scenario describes, for instance, an NV ensemble  used as a detector for a $^1$H spin cluster out of the diamond sample~\cite{Aslam17}. As in the previous case, the ideal solid-blue curve in Fig.~\ref{results} (d) has been obtained by delivering instantaneous $\pi$ pulses, and in absence of control errors. In the same figure, the squares represent the signal obtained with our method, i.e. by using  the controls in Fig.~\ref{results} (e) (an animation of the NV spin state evolution during the $\pi$ pulse is available in~\cite{SM}) and averaging the responses of of 10 NVs where the detuning error has been randomly taken from a Gaussian distribution centered at $\xi_\delta = 0$ and with a width of 1 MHz. More specifically, we have used the following values $\xi_\delta = (2\pi)\times[ 0.5376, 1.8338, -2.2588, 0.8622, 0.3188, -1.3076,\\ -0.4336, 0.3426, -2.7784, 2.1694]$ MHz, while the Rabi frequency deviation is $\xi_\Omega = 1\%$ for all cases. One can observe that this average spectrum fully overlaps with the ideal NV response, which demonstrates the good performance of our method. The circles in Fig.~\ref{results} (d) denotes the signal obtained with top-hat $\pi$ pulses with a Rabi frequency  $\Omega_{\rm th} = (2\pi)\times 40$ MHz, i.e. equal to the maximum amplitude of $\Omega(t)$ in Fig.~\ref{results} (e).  Again, the signal-contrast produced by standard top-hat $\pi$ pulses is much lower than the one achieved by our method which further confirm the advantages of the latter. Finally, in Fig.~\ref{results} (f) we plot the average signal obtained with the $\pi$ pulses in Ref.~\cite{Casanova18}, and for the same errors in Fig.~\ref{results} (d). We can observe that the spectrum in Fig.~\ref{results} (f) cannot  offer any information regarding the scanned sample while, with our method, we can clearly observe a resonance peak that meets the ideal response leading to reliable identification.  

\section{Conclusions} We have demonstrated that the integration of STA techniques in the design of DD sequences leads to superior performance in the detection of high frequency target signals. Our method exhibit an enhanced resilience against typical control errors, and can be straightforwardly incorporated to any DD sequence used in Nanoscale NMR. We exemplified our theory in the frame of Nanoscale NMR with NV centers. However, our method is general and applicable to other solid-state quantum sensor devices such as silicon vacancy centers, germanium vacancy centers, or divacancies in silicon carbide.

\begin{acknowledgements}
We acknowledge financial support from Spanish Government via PGC2018-095113-B-I00 and EUR2020-112117 (MCIU/AEI/FEDER, UE), Basque Government via IT986-16, as well as from QMiCS
(820505) and OpenSuperQ (820363) of the EU Flagship on Quantum Technologies, and the EU FET Open Grant Quromorphic (828826). J. C. acknowledges the Ram\'{o}n y Cajal program (RYC2018-025197-I) and the EUR2020-112117 project of the Spanish MICIN, as well as support from the UPV/EHU through the grant EHUrOPE. X. C. acknowledges NSFC (11474193), SMSTC (2019SHZDZX01-ZX04, 18010500400 and 18ZR1415500), the Program for Eastern Scholar and the Ram\'{o}n y Cajal program (RYC2017-22482). 
\end{acknowledgements}

\section*{Appendix A: The coupling condition}

\begin{figure*}[t]
\hspace{-0. cm}\includegraphics[width=1.4\columnwidth]{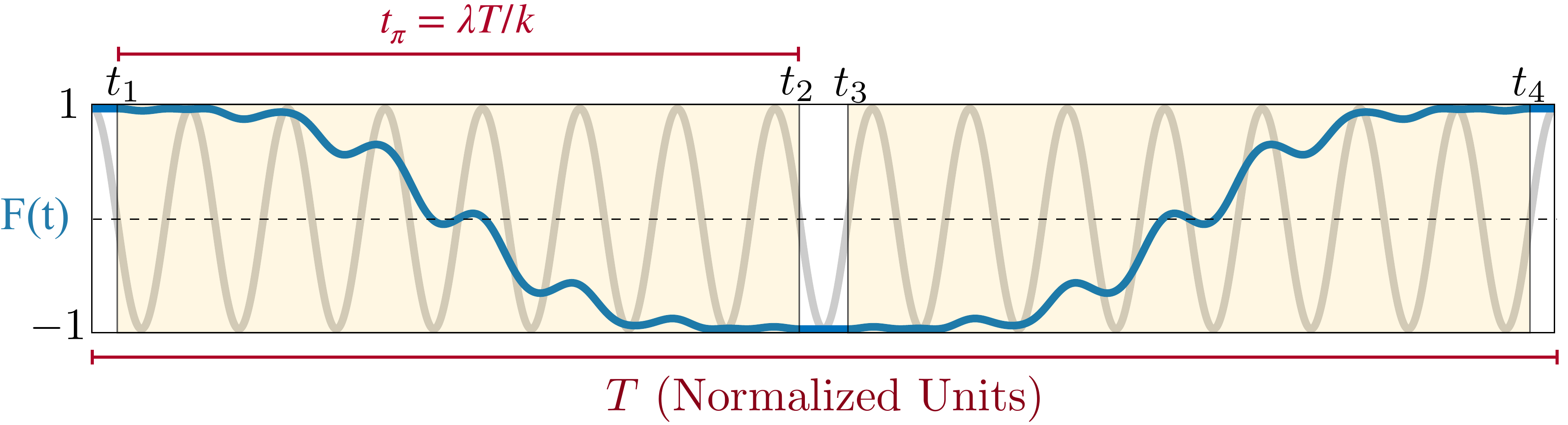}
\caption{Modulation function $F(t)$ corresponding to a $\pi$ pulse generated using our method (blue line) and the distribution of times for $\pi$ pulses. These are displayed in the yellow areas.}
\label{Spulsetimes}
\end{figure*}

In this section we derive the {\it coupling condition} in Eq.~(8) of the main text. We start from the Hamiltonian in Eq.~(2) where, for the sake of simplicity, we consider a target classical signal. This is 

 \begin{equation}
 H =  \frac{\Gamma}{2} \sigma_z \cos(\omega_s t)+ \frac{\Omega(t)}{2}\sigma_{\phi}+\sigma_z\frac{\delta(t)}{2}
\end{equation}
In the rotating frame of the control (i.e. of $\frac{\Omega(t)}{2}\sigma_{\phi}+\sigma_z\frac{\delta(t)}{2}$) this Hamiltonian is
\begin{equation}
H_I = F(t)\frac{\Gamma}{2}\sigma_z \cos(\omega_s t),
\end{equation}
with $F(t)$ being a modulation function that appears due to the action of the control on the NV $\sigma_z$ operator. In particular $F(t) = \pm 1$ in the regions where the controls are switched off, this is in the regions without $\pi$ pulses, while $F(t)$ adopts specific forms during $\pi$ pulse execution depending on the value of the controls $\Omega(t)$ and $\delta(t)$, see Fig.~\ref{Spulsetimes}. If we consider a pulse sequence of period $T$, the modulation function $F(t)$ can be expanded using Fourier series as $F(t) = \sum_k f_k \cos(k\omega_m t)$, with $\omega_m = \frac{2\pi}{T}$ and $k$ being a natural odd number. In particular, we can compute $f_k$ as
 \begin{equation}
 \begin{split}
 f_k = & \frac{2}{T}\left[\int_{0}^{t_1} \cos(k \omega_m t) dt+\int_{t_1}^{t_2} F(t) \cos(k \omega_m t) dt- \right. \\
 & - \int_{t_2}^{t_3} \cos(k \omega_m t) dt+\int_{t_3}^{t_4} F(t)\cos(k \omega_m t) dt+ \\ 
 & + \left. \int_{t_4}^{T} \cos(k \omega_m t) dt\right].
 \end{split}
 \end{equation}
 Here, the distribution of times $t_{i}$ (for $i=1,2,3,4$) can be seen in Fig.~\ref{Spulsetimes}. In particular, in this figure  we consider a situation where the harmonic $k=15$ is used to carry on the NV-nucleus coupling, and two $\pi$ pulses are displayed with a length including 7 oscillations of the function $\cos(k \omega_m t)$ (i.e. $\lambda=7$).

If the function $F(t)$ is designed such that $\int_{t_1}^{t_2} F(t) \cos(k \omega_m t) dt = \int_{t_3}^{t_4} F(t)\cos(k \omega_m t) dt =0$, we get
\begin{equation}
\begin{split}
f_k = &\frac{2}{T}\left[\int_{0}^{t_1} \cos(k \omega_m t) dt-\int_{t_2}^{t_3} \cos(k \omega_m t) dt + \right. \\
&\left. +\int_{t_4}^{T} \cos(k \omega_m t) dt\right]= \frac{4}{k\pi}\sin\left(\frac{k\pi}{T}\right)\cos\left(\frac{k\pi}{T}t_\pi\right).
\end{split}
\end{equation}
It can be seen then that the optimal value $|f_k| = \frac{4}{k\pi}$ (this is the value for $f_k$ that one would get if applies instantaneous $\pi$ pulses) is obtained when $t_\pi = \frac{\lambda T}{k}$ with $\lambda$ being a natural number. 

Then, to achieve maximal coupling (i.e. $|f_k| = \frac{4}{k\pi}$) it is mandatory to nullify the integrals during the $\pi$ pulses. As previously mentioned, the modulation function appears owing to the action of the controls on the $\sigma_z$ NV operator. In particular, and calling $U_0(t) = \hat T \exp{\left[-i\int_{t_0}^t  \frac{\Omega(s)}{2}\sigma_{\phi}+\sigma_z\frac{\delta(s)}{2} ds\right]}$ we have that
\begin{equation}
U_0(t)^\dag \sigma_z U_0(t) = F(t)\sigma_z+F_x(t)\sigma_x+F_y(t)\sigma_y .
\end{equation}
If a sequence with alternating pulses is employed, note this is our case as we use the XY8 sequence, the $F_x(t)$, and $F_y(t)$ functions do not have an effect at the resonance point (this is when $k\omega_m = \omega_s$).  Then, the $F_{x,y}$ components can be neglected.

Now we can write  (assuming that $|\phi(0)\rangle = |1\rangle$)
\begin{equation}
\langle\phi(0)|U_0^{\dag}(t)\sigma_z U_0(t)|\phi(0)\rangle = \langle\phi(t)|\sigma_z |\phi(t)\rangle= F(t).
\end{equation}
Finally, using the expression for $|\phi(t)\rangle$ in Eq.~(3) we can compute that $F(t)=\cos(\theta)$, which leads to the coupling condition in Eq.~(8) of the main text
\begin{equation}
\int_{0}^{t_\pi}\cos (\theta) \cos(k \omega_m t) dt =0.
\end{equation}
\section*{Appendix B: Error cancelation condition}
Here we show the derivation of the approximate transition probability in the presence of errors. We start from the control Hamiltonian including errors. This is
\begin{equation}
H_c + H_\epsilon = \frac{\Omega(t)(1+\xi_\Omega)}{2}\sigma_{\phi}+\frac{\delta(t)+\xi_\delta}{2}\sigma_z.
\end{equation} 
Now we move to a rotating frame w.r.t. the control. This leads to 
\begin{equation}
H_I = U_0^{\dag}(t)H_\epsilon U_0(t),\label{Interaction Picture}
\end{equation}
where $U_0(t) = \hat T \exp{\left[-i\int_{t_0}^t  \frac{\Omega(s)}{2}\sigma_{\phi}+\sigma_z\frac{\delta(s)}{2} ds\right]}$ is the control Hamiltonian propagator. We can expand now the interaction picture propagator using Dyson series
\begin{equation}
U_I(t_\pi, 0) = \mathbb{I}-i\int_{0}^{t_\pi}dt H_I(t)-\int_{0}^{t_\pi}dt \int_{0}^{t}dt' H_I(t)H_I(t')+...
\end{equation}
If we now write $H_I(t)$ as in Eq.~(\ref{Interaction Picture}), and multiply the previous expression by $|\phi(0)\rangle$ we get  (up to the second order)
\begin{equation}
\begin{split}
|\phi(t_\pi)\rangle_I \approx \ &|\phi_0(0)\rangle-i\int_{0}^{t_\pi}dt H_I (t) |\phi_0(0)\rangle_I-\\
&-\int_{0}^{t_\pi}dt \int_{0}^{t}dt' H_I(t)H_I(t') |\phi_0(0)\rangle_I,
\end{split}
\end{equation}
where $|\cdot\rangle_I$ represents the state in the interaction picture, while $|\phi_0(t)\rangle$ is the state evolved without errors.  

Now we apply $U_0(t_\pi, 0)$ to $|\phi(t_\pi)\rangle_I$ and find
\begin{equation}
\begin{split}
|\phi(t_\pi)\rangle \approx \ & |\phi_0(t_\pi)\rangle-i\int_{0}^{t_\pi}dt U_0 (t_\pi, t)H_\epsilon |\phi_0(t)\rangle-\\
&-\int_{0}^{t_\pi}dt \int_{0}^{t}dt' U_0(t_\pi, t) H_\epsilon U_0 (t, t')H_\epsilon |\phi_0(t')\rangle.
\end{split}
\end{equation}

At this point we make use of the relation $U_0(t_f, 0) = |\phi_0(t_f)\rangle\langle\phi_0(0)|+|\phi_0^\perp(t_f)\rangle\langle\phi_0^\perp(0)|$, where $|\phi_0^\perp(t)\rangle  = \left[ \sin \left(\frac{\theta}{2}\right)e^{i\frac{\beta}{2}} |1\rangle -\cos\left(\frac{\theta}{2}\right)  e^{-i\frac{\beta}{2}} |0\rangle \right] e^{-i \gamma}$ is the orthogonal state to $|\phi (t) \rangle = \left[\cos\left(\frac{\theta}{2}\right) e^{i\frac{\beta}{2}} |1\rangle + \sin \left(\frac{\theta}{2}\right) e^{-i\frac{\beta}{2}} |0\rangle \right] e^{i \gamma}$. Writing the full expression of $H_\epsilon$ and using the identity $\int_{a}^{b}dx\int_{a}^x dy f(x, y) = \frac{1}{2}\int_{a}^{b}dx\int_{a}^{b} dy f(x, y)$ if $f(x, y) = f(y, x)$ in the integration range, we can obtain that the probability to find $|\phi_0(t_\pi)\rangle$ at the end of the pulse up to order two, i.e. $P(t_\pi) = |\langle\phi_0(t_\pi)|\phi(t_\pi)\rangle|^2$, is
\begin{equation}
P(t_\pi) \approx 1 - \left|\int_{0}^{t_\pi}dt\langle\phi_0^\perp(t)|\left(\frac{\xi_\delta}{2}\sigma_z+\frac{\Omega\ \xi_\Omega}{2}\sigma_\phi\right)|\phi_0(t)\rangle\right|^2.
\end{equation}
Finally, using expression (4) we get the {\it error cancelation condition}
\begin{equation}
P(t_\pi) \approx 1-\left|\int_{0}^{t_\pi}dt \frac{e^{i2 \gamma(t)}}{2}\left(\xi_\delta\sin(\theta) - i2\xi_\Omega\dot{\theta}\sin^2(\theta)\right)\right|^2\ .
\end{equation}


\begin{thebibliography}{99} 
\bibitem{Mamin13} H. J. Mamin, M. Kim, M. H. Sherwood, C. T. Rettner, K. Ohno, D. D. Awschalom, and D. Rugar, Nanoscale Nuclear Magnetic Resonance with a Nitrogen-Vacancy Spin Sensor, Science {\bf 339}, 557 (2013).
\bibitem{Muller14} C.  M{\"u}ller, X. Kong, J.-M. Cai, K. Melentijevi\'c, A.Stacey, M. Markham, D. Twitchen, J. Isoya, S. Pezzagna, J. Meijer, J. F. Du, M. B. Plenio, B. Naydenov, L. P. McGuinness, and F. Jelezko, Nuclear magnetic resonance spectroscopy with single spin sensitivity, Nat. Commun. {\bf 5}, 4703 (2014).
\bibitem{DeVience15} S. J. DeVience, L. M. Pham, I. Lovchinsky, A. O. Sushkov, N. Bar-Gill, C. Belthangady, F. Casola, M. Corbett, H. Zhang, M. Lukin, H. Park, A. Yacoby, and R. L. Walsworth, Nanoscale NMR spectroscopy and imaging of multiple nuclear species, Nature Nanotech {\bf 10}, 129 (2015).
\bibitem{Degen17} C. L. Degen, F. Reinhard, and P. Cappellaro, Quantum sensing, Rev. Mod. Phys. {\bf 89}, 035002 (2017).
\bibitem{Schwartz19} I. Schwartz, J. Rosskopf, S. Schmitt, B. Tratzmiller, Q. Chen, L. P. McGuinness, F. Jelezko, and Martin B. Plenio, Blueprint for nanoscale NMR, Sci. Rep. {\bf 9}, 6938 (2019).
\bibitem{Zopes18} J. Zopes, K. Herb, K. S. Cujia, and C. L. Degen, Three-Dimensional Nuclear Spin Positioning Using Coherent Radio-Frequency Control, Phys. Rev. Lett. {\bf 121}, 170801 (2018).
\bibitem{Zopes18bis} J. Zopes, K. S. Cujia, K. Sasaki, J. M. Boss, K. M. Itoh, and C. L. Degen, Three-dimensional localization spectroscopy of individual nuclear spins with sub-Angstrom resolution, Nat. Commun. {\bf 9}, 4678 (2018).
\bibitem{Bradley19}C. E. Bradley, J. Randall, M. H. Abobeih, R. C. Berrevoets, M. J. Degen, M. A. Bakker, M. Markham, D. J. Twitchen, and T. H. Taminiau, A Ten-Qubit Solid-State Spin Register with Quantum Memory up to One Minute, Phys. Rev. X {\bf 9}, 031045 (2019).
\bibitem{Abobeih19} M. H. Abobeih, J. Randall, C. E. Bradley, H. P. Bartling, M. A. Bakker, M. J. Degen, M. Markham, D. J. Twitchen, and T. H. Taminiau, Atomic-scale imaging of a 27-nuclear-spin cluster using a quantum sensor, Nature {\bf 576}, 411 (2019).
\bibitem{Holzgrafe20} J. Holzgrafe, Q. Gu, J. Beitner, D. M. Kara, H. S. Knowles, and Mete Atat{\"u}re, Nanoscale NMR Spectroscopy Using Nanodiamond Quantum Sensors, Phys. Rev. Applied {\bf 13}, 044004 (2020).
\bibitem{Schmitt17}S. Schmitt, T. Gefen, F. M. St{\"u}rner, T. Unden, G. Wolff, C. M{\"u}ller, J. Scheuer, B. Naydenov, M. Markham, S. Pezzagna, J. Meijer, I. Schwarz, M. Plenio, A. Retzker, L. P. McGuinness, and F. Jelezko, Submillihertz magnetic spectroscopy performed with a nanoscale quantum sensor, Science {\bf 356}, 832 (2017).
\bibitem{Boss17} J. M. Boss, K. S. Cujia, J. Zopes, and C. L. Degen, Quantum sensing with arbitrary frequency resolution, Science {\bf 356}, 837 (2017).
\bibitem{Glenn18} D. R. Glenn, D. B. Bucher, J. Lee, M. D. Lukin, H. Park, and R. L. Walsworth, High-resolution magnetic resonance spectroscopy using a solid-state spin sensor, Nature {\bf 555}, 351 (2018).
\bibitem{Staudacher13}T. Staudacher, F. Shi, S. Pezzagna, J. Meijer, J. Du, C. A. Meriles, F. Reinhard, and  J. Wrachtrup, Nuclear Magnetic Resonance Spectroscopy on a (5-Nanometer)$^3$ Sample Volume, Science {\bf 339}, 561 (2013).
\bibitem{Shi15} F. Shi, Q. Zhang, P. Wang, H. Sun, J. Wang, X. Rong, M. Chen, C. Ju, F. Reinhard, H. Chen, J. Wrachtrup, J. Wang, and J. Du, Single-protein spin resonance spectroscopy under ambient conditions, Science {\bf 347}, 1135 (2015).
\bibitem{Lovchinsky16} I. Lovchinsky, A. O. Sushkov, E. Urbach, N. P. de Leon, S. Choi, K. De Greve, R. Evans, R. Gertner, E. Bersin, C. M{\"u}ller, L. McGuinness, F. Jelezko, R. L. Walsworth, H. Park, and M. D. Lukin, Nuclear magnetic resonance detection and spectroscopy of single proteins using quantum logic, Science {\bf 351}, 836 (2016).
\bibitem{Aslam17} N. Aslam, M. Pfender, P. Neumann, R. Reuter, A. Zappe, F. F. de Oliveira, A. Denisenko, H. Sumiya, S. Onoda, J. Isoya, and J. Wrachtrup, Nanoscale nuclear magnetic resonance with chemical resolution, Science {\bf 357}, 67 (2017).
\bibitem{Kucsko13} G. Kucsko, P. C. Maurer, N. Y. Yao, M. Kubo, H. J. Noh, P. K. Lo, H. Park, and M. D. Lukin, Nanometre-scale thermometry in a living cell, Nature {\bf 500}, 54 (2013).
\bibitem{Choi20} J. Choi, H. Zhou, R. Landig, H.-Y. Wu, X. Yu, S. Von Stetina, G. Kucsko, S. Mango, D. Needleman, A. D. T. Samuel, P. Maurer, H. Park, M. D. Lukin, Probing and manipulating embryogenesis via nanoscale thermometry and temperature control, PNAS {\bf 117}, 14636 (2020).
\bibitem{Glover02} P. Glover and P. Mansfield, Limits to magnetic resonance microscopy, Rep. Prog. Phys. {\bf 65}, 1489 (2002).
\bibitem{Rogers14} L. J. Rogers, K. D. Jahnke, M. H. Metsch, A. Sipahigil, J. M. Binder, T. Teraji, H. Sumiya, J. Isoya, M. D. Lukin, P. Hemmer, and F. Jelezko, All-Optical Initialization, Readout, and Coherent Preparation of Single Silicon-Vacancy Spins in Diamond, Phys. Rev. Lett. {\bf 113}, 263602 (2014).
\bibitem{Christle17} D. J. Christle, P. V. Klimov, C. F. de las Casas, K. Sz\'asz, V. Iv\'{a}dy, V. Jokubavicius, J. Ul Hassan, M. Syv{\"a}j{\"a}rvi, W. F. Koehl, T. Ohshima, N. T. Son, E. Janz\'{e}n, A. Gali, and D. D. Awschalom, Isolated Spin Qubits in SiC with a High-Fidelity Infrared Spin-to-Photon Interface, Phys. Rev. X {\bf 7}, 021046 (2017).
\bibitem{Siyushev17} P. Siyushev, M. H. Metsch, A. Ijaz, J. M. Binder, M. K. Bhaskar, D. D. Sukachev, A. Sipahigil, R. E. Evans, C. T. Nguyen, M. D. Lukin, P. R. Hemmer, Y. N. Palyanov, I. N. Kupriyanov, Y. M. Borzdov, L. J. Rogers, and Fedor Jelezko, Optical and microwave control of germanium-vacancy center spins in diamond, Phys. Rev. B {\bf 96}, 081201(R) (2017).
\bibitem{Doherty13} M. W. Doherty, N. B. Manson, P. Delaney, F. Jelezko, J. Wrachtrup, and  L. C. L. Hollenberg, The nitrogen-vacancy colour centre in diamond, Phys. Rep. {\bf 528}, 1 (2013). 
\bibitem{Dobrovitski13} V. V. Dobrovitski, G. D. Fuchs, A. L. Falk, C. Santori, and D.D. Awschalom, Quantum Control over Single Spins in Diamond, Annu. Rev. Condens. Matter Phys. {\bf 4}, 23 (2013).
\bibitem{Rondin14} L. Rondin, J. P. Tetienne, T. Hingant, J. F. Roch, P. Maletinsky, and V. Jacques, Magnetometry with nitrogen-vacancy defects in diamond, Rep. Prog. Phys. {\bf 77}, 056503 (2014).
\bibitem{Schirhagl14} R. Schirhagl, K. Chang, M. Loretz, and C. L. Degen, Nitrogen-Vacancy Centers in Diamond: Nanoscale Sensors for Physics and Biology, Annu. Rev. Phys. Chem. {\bf 65}, 83 (2014).
\bibitem{Wu16} Y. Wu, F. Jelezko, M. B. Plenio, and T. Weil, Diamond Quantum Devices in Biology, Angew. Chem. {\bf 55}, 6586 (2016).
\bibitem{Balasubramanian08} G. Balasubramanian, I. Y. Chan, R. Kolesov, M. Al-Hmoud, J. Tisler, C. Shin, C. Kim, A. Wojcik, P. R. Hemmer, A. Krueger, T. Hanke, A. Leitenstorfer, R. Bratschitsch, F. Jelezko, and  J. Wrachtrup, Nanoscale imaging magnetometry with diamond spins under ambient conditions, Nature {\bf 455}, 648 (2008).
\bibitem{Pham16} L. M. Pham, S. J. DeVience, F. Casola, I. Lovchinsky, A. O. Sushkov, E. Bersin, J. Lee, E. Urbach, P. Cappellaro, H. Park, A. Yacoby, M. Lukin, and R. L. Walsworth, NMR technique for determining the depth of shallow nitrogen-vacancy centers in diamond, Phys. Rev. B {\bf 93}, 045425 (2016).
\bibitem{Kehayias17} P. Kehayias, A. Jarmola, N. Mosavian, I. Fescenko, F. M. Benito, A. Laraoui, J. Smits, L. Bougas, D. Budker, A. Neumann, S. R. J. Brueck, and V. M. Acosta, Solution nuclear magnetic resonance spectroscopy on a nanostructured diamond chip, Nature Communications {\bf 8}, 188 (2017).
\bibitem{Chipaux18} M. Chipaux, K. J. van der Laan, S. R. Hemelaar, M. Hasani, T. Zheng, and R. Schirhagl, Nanodiamonds and Their Applications in Cells, Small {\bf 14}, 1704263 (2018).
\bibitem{Maudsley86} A. A. Maudsley, Modified Carr-Purcell-Meiboom-Gill sequence for NMR fourier imaging applications, J. Magn. Reson. {\bf 69}, 488 (1986).
\bibitem{Uhrig08} G. Uhrig, Exact results on dynamical decoupling by $\pi$ pulses in quantum information processes, New J. Phys. {\bf 10}, 083024 (2008).
\bibitem{Pasini08} S. Pasini, T. Fischer, P. Karbach, and G. S. Uhrig, Optimization of short coherent control pulses, Phys. Rev. A {\bf 77}, 032315 (2008).
\bibitem{Souza11}A. M. Souza, G. A. \'Alvarez, and D. Suter, Robust Dynamical Decoupling for Quantum Computing and Quantum Memory, Phys. Rev. Lett. {\bf 106}, 240501 (2011).
\bibitem{Wang11}Z. Y. Wang and R.-B. Liu, Protection of quantum systems by nested dynamical decoupling, Phys. Rev. A 83, 022306 (2011).
\bibitem{Souza12} A. M. Souza, G. A. \'Alvarez, and D. Suter, Robust dynamical decoupling, Phil. Trans. R. Soc. A {\bf 370}, 4748 (2012).
\bibitem{Casanova15} J. Casanova, Z. Y. Wang, J. F. Haase, and M. B. Plenio, Robust dynamical decoupling sequences for individual-nuclear-spin addressing, Phys. Rev. A {\bf 92}, 042304 (2015).
\bibitem{Wang16} Z. Y. Wang, J. F. Haase, J. Casanova, and M. B. Plenio, Positioning nuclear spins in interacting clusters for quantum technologies and bioimaging, Phys. Rev. B {\bf 93}, 174104 (2016).
\bibitem{Lang17} J. E. Lang, J. Casanova, Z. Y. Wang, M. B. Plenio, and T. S. Monteiro, Enhanced Resolution in Nanoscale NMR via Quantum Sensing with Pulses of Finite Duration, Phys. Rev. Applied {\bf 7}, 054009 (2017). 
\bibitem{Wang19} Z. Y. Wang, J. E. Lang, S. Schmitt, J. Lang, J. Casanova,  L. McGuinness, T. S. Monteiro, F. Jelezko, and M. B. Plenio, Randomization of Pulse Phases for Unambiguous and Robust Quantum Sensing, Phys. Rev. Lett. {\bf 122}, 200403 (2019).
\bibitem{Hirose12} M. Hirose, C. D. Aiello, and P. Cappellaro, Continuous dynamical decoupling magnetometry, Phys. Rev. A {\bf 86}, 062320 (2012).
\bibitem{Cai13} J. M. Cai, F. Jelezko, M. B. Plenio, and A. Retzker, Diamond-based single-molecule magnetic resonance spectroscopy, New J. Phys. {\bf 15}, 013020 (2013).
\bibitem{Puebla18} R. Puebla, J. Casanova, and M. B. Plenio, A robust scheme for the implementation of the quantum Rabi model in trapped ions, New J. Phys. {\bf 18}, 113039 (2016).
\bibitem{Arrazola19} I. Arrazola, M. B. Plenio, E. Solano, and J. Casanova, Hybrid Microwave-Radiation Patterns for High-Fidelity Quantum Gates with Trapped Ions, Phys. Rev. Applied {\bf 13}, 024068 (2020).
\bibitem{Levitt08} M. H. Levitt, \textit{Spin Dynamics: Basics of Nuclear Magnetic Resonance} (Wiley, West Sussex, 2008).
\bibitem{Reynhardt01} E. C. Reynhardt and G. L. High, Nuclear magnetic resonance studies of diamond, Prog. Nucl. Magn. Reson. Spectrosc. {\bf 38}, 37 (2001).
\bibitem{Hartmann62} S. R. Hartmann and E. L. Hahn, Nuclear Double Resonance in the Rotating Frame, Phys. Rev. {\bf 128}, 2042 (1962).
\bibitem{Casanova18} J. Casanova, Z. Y. Wang, I. Schwartz, and M. B. Plenio, Shaped Pulses for Energy-Efficient High-Field NMR at the Nanoscale, Phys. Rev. Applied {\bf 10}, 044072 (2018).
\bibitem{Genov19} G. T. Genov, Y. Ben-Shalom, F. Jelezko, A. Retzker and N. Bar-Gill, Efficient and robust signal sensing by sequences of adiabatic chirped pulses, Phys. Rev. Research. {\bf 2}, 033216 (2020).
\bibitem{Erik13} E. Torrontegui, S. Ib\'anez, S. Mart\'inez-Garaot, M. Modugno, A. del Campo, D. Gu\'ery-Odelin, A. Ruschhaupt, X. Chen, and J. G. Muga, Shortcuts to adiabaticity, Adv. At. Mol. Opt. Phys. {\bf 62}, 117 (2013).
\bibitem{Guery19} D. Gu\'ery-Odelin, A. Ruschhaupt, A. Kiely, E. Torrontegui, S. Mart\'inez-Garaot, and J. G. Muga, Shortcuts to adiabaticity: Concepts, methods, and applications, Rev. Mod. Phys. {\bf 91}, 045001 (2019).
\bibitem{Laraoui11} A. Laraoui, J. S. Hodges, C. A. Ryan, and C. A. Meriles, Diamond nitrogen-vacancy center as a probe of random fluctuations in a nuclear spin ensemble, Phys. Rev. B {\bf 84}, 104301 (2011).
\bibitem{Daems13} D. Daems, A. Ruschhaupt, D. Sugny, and S. Gu\'erin, Robust Quantum Control by a Single-Shot Shaped Pulse, Phys. Rev. Lett. {\bf 111}, 050404 (2013).
\bibitem{Ruschhaupt12} A. Ruschhaupt, X. Chen, D. Alonso, and J. G. Muga, Optimally robust shortcuts to population inversion in two-level quantum systems, New J. Phys. {\bf 14}, 093040 (2012). 
\bibitem{Lu13} X.-J. Lu, X. Chen, A. Ruschhaupt, D. Alonso, S. Gu\'erin, and J. G. Muga, Fast and robust population transfer in two-level quantum systems with dephasing noise and/or systematic frequency errors, Phys. Rev. A {\bf 88}, 033406 (2013).
\bibitem{Qi17} Q. Zhang, X. Chen, and D. Gu\'ery-Odelin, Reverse engineering protocols for controlling spin dynamics, Sci. Rep. {\bf 7}, 15814 (2017). 
\bibitem{Blackman58} R. B. Blackman and J. W. Tukey, in \textit{The Measurement of Power Spectra from the Point of View of Communications Engineering} (Dover Publications Inc., New York, 1958), p. 221.
\bibitem{SM} See Supplemental Material at [URL will be inserted by publisher] for [two animations of the NV trajectories in the Bloch sphere induced by our controls].
\end{thebibliography}
\end{document}